%% file: ARXIV_main.tex
\documentclass{article}
\pdfoutput=1
\usepackage{url}
\usepackage{hyperref}
\usepackage{colortbl}
\usepackage{xcolor}

\usepackage{geometry}
\def\UrlBreaks{\do\/\do-}

\title{Open Data Resources for Fighting COVID-19}
\author{Teodoro Alamo\footnote{Departamento de Ingenier\'ia de Sistemas y Autom\'atica, Universidad de Sevilla, Escuela Superior de Ingenieros, Camino de los Descubrimientos s/n, 41092 Sevilla, Spain (e-mail: talamo@us.es) }, Daniel G. Reina\footnote{Departamento de Ingenier\'ia Electrónica, Universidad de Sevilla, Escuela Superior de Ingenieros, Camino de los Descubrimientos s/n, 41092 Sevilla, Spain (e-mail: dgutierrezreina@us.es) }, Martina Mammarella\footnote{{Institute of Electronics, Computer and Telecommunication Engineering, National Research Council of Italy, Turin, Italy (e-mail: martina.mammarella@ieiit.cnr.it).}}, Alberto Abella\footnote{FIWARE Foundation. Germany (e-mail: alberto.abella@gmail.com)}}

\begin{document}

\maketitle
\begin{abstract}
We provide an insight into the open data resources pertinent to the study of the spread of Covid-19 pandemic and its control. We identify the variables required to analyze fundamental aspects like seasonal behaviour, regional mortality rates, and effectiveness of government measures. Open data resources, along with data-driven methodologies, provide many opportunities to improve the response of the different administrations to the virus. We describe the present limitations and difficulties encountered in most of the open data resources. To facilitate the access to the main open data portals and resources, we identify the most relevant institutions, at a world scale, providing Covid-19 information and/or auxiliary variables (demographics, mobility, etc.). We also describe several open resources to access Covid-19 datasets at a country-wide level (i.e., China, Italy, Spain, France, Germany, U.S., etc.). In an attempt to facilitate the rapid response to the study of the seasonal behaviour of Covid-19, we enumerate the main open resources in terms of weather and climate variables. We also assess the reusability of some representative open data sources.
\end{abstract}

\footnote{This research was funded by "Plan Propio de la Universidad de Sevilla" under the contract of "Contratos de acceso al Sistema Español de Ciencia, Tecnología e Innovación para el desarrollo del programa propio de I+D+i de la Universidad de Sevilla" of Dr. Daniel G. Reina.}
\newpage

\tableofcontents

\newpage
\input{Introduction}
\input{Covid_19}
\input{Data_driven_ways_to_fight_the_pandemic}
\input{Limitations_and_challenges_raised_by_the_available_data}
\input{Open_data_institutions_providing_worldwide_Covid_19_data}

\input{Open_source_communities}
\input{Covid_19_Data_Sets}
\input{Data_sets_of_relevant_variables_for_Covid_19_analysis}
\input{Reusability_of_Open_Data_Sources}
\input{Conclusions_journal}

\input{Acknowledgments}

\bibliography{Bib_Covid_19}

\end{document}

%% file: Introduction.tex
\section{Introduction}
We provide in this document a survey on the main open-resources for addressing the Covid-19 pandemic from a data science point of view. Since the number of institutions and research teams working nowadays against the virus is growing at a very fast pace, it is impossible to provide an exhaustive list of all the meaningful open data providers. At a global world scope, we identify the most relevant sources. However, the enumeration of the regional institutions providing local information is so extensive that we address it specifically only for some countries (like China, Italy, Spain, and the U.S, among others). We focus on the variables that have possible effects on the evolution and control of the disease at a global and regional scale \cite{lakshmi2020factors}. That is, we do not cover in this document the data specifically related to medical treatments, vaccines, etc. \cite{le2020covid}. We do provide open resources for the number of hospitalized cases, intensive care units (ICU) cases, number of tests, etc. These variables are very relevant to monitor the evolution of the pandemic and also to evaluate the actions taken by the decision-makers \cite{lakshmi2020factors}. 

With this document, we try to make accessible many significant open data resources on Covid-19 for the scientific community. In many situations, identifying adequate sources is difficult, especially for non-expert data scientists. For example, GitHub repository contains many meaningful datasets of global and regional scope, but it might be challenging to discover them without adequate guidance. Besides, the reliability of the data source provider can be a concern. Therefore, this paper is aimed at providing a big picture of the available data source providers for analyzing Covid-19 propagation and control. We have tried to find stable and reliable resources so that the utility of this paper endures in time.

The paper is organized as follows. We first analyze in Section \ref{sec:Covid19} the different variables that have a significant effect on the evolution and control of the epidemic (demographics, mobility, weather conditions, government measures, etc.). The opportunities that open data resources on Covid-19 offer to fight the pandemic are highlighted, from a data-driven perspective, in Section \ref{sec:Data:Driven}. 
Different limitations and inaccuracies of the currently available sources, along with the difficulties encountered when using them in a data-science context are discussed in Section \ref{Sec:Limitations}. The most relevant open data institutions at a global scale, addressing the Covid-19 pandemic, are enumerated in Section \ref{sec:Global:OpenDataInstitutions}. More functionally, in Section \ref{sec:Open:Data:Repositories}, we identify open source communities that facilitate access to the required data. In Section \ref{sec:Covid19:DataSets}, we identify open datasets related to specific Covid-19 variables at a global and regional scale. The open access to auxiliary variables of interest to model specific aspects of the pandemic, like seasonal behaviour or local mortality rate, is described in Section \ref{Sec:Relevant:Variables}. In Section \ref{sec:Reusability}, we discuss the reusability of the available datasets. Finally, a concluding Section \ref{sec:conclusions} is included. 

%% file: Covid_19.tex
\section{Covid-19}\label{sec:Covid19}

Coronavirus disease 2019 (Covid-19), technically known as SARS-CoV-2, is an infectious disease that was first identified on December 31st 2019 in Wuhan, the capital of China's Hubei province. The World Health Organisation (WHO) declared the 2019–20 coronavirus outbreak a Public Health Emergency of International Concern on January 30th 2020 and a pandemic on March 11th.
 
 
The virus is mainly spread during close contact and by small droplets produced when those infected cough, sneeze or talk. These small droplets may also be produced during breathing. The virus is most contagious during the first 4-6 days  after onset of symptoms \cite{Ferretti2020}, although spread is possible in asymptomatic conditions \cite{YanBai2020} and in later stages of the disease \cite{Ferretti2020}. The time from exposure to onset of symptoms (incubation period) is typically around 5 days but may range from 2 to 14 days \cite{lauer2020incubation}. Recommended measures to control the pandemic include social distancing, mobility constraints, pro-active testing and isolation of detected cases \cite{Hellewell2020}.  

\subsection{Covid-19 cases in the world}
To monitor the spread of Covid-19, the different regional institutions are measuring the number of confirmed cases, deaths, recovered, hospitalized cases, intensive care unit (ICU) cases, etc. Because of the incubation period \cite{lauer2020incubation}, all these variables are related with the number of infected cases in a \textit{delayed} way. One of the main objectives of those institutes is to estimate the basic reproductive number $R_{0}$, which serves to characterize the spread of the virus \cite{Martcheva2015}. Several works have calculated $R_{0}$ for some outbreaks of specific locations. The estimated values are ranging from 2 to 3 \cite{liu2020reproductive}. However, only limited data have been used in the majority of works. On the other hand, achieving an accurate model of the virus reproduction is a challenging task, which involves many variables and validation steps. Unfortunately, the open data sets available nowadays are locally collected, imprecise with different criteria (lack of standardization on data collection), inconsistent with data models, and incomplete.

One of the main limitations of these datasets is that often only cases confirmed by a laboratory test are included. The standard method of diagnosis is by real-time reverse transcription polymerase chain reaction (RT-PCR) from a nasopharyngeal swab. The infection can also be diagnosed from a combination of symptoms, risk factors and a chest CT (computed tomography) scan showing features of pneumonia. Thus, on a general basis, the infected cases without a positive laboratory test are not considered confirmed cases in the time-series data available on the different open-source repositories. The same problem can be encountered when analyzing death cases. In many situations, specially at the beginning of the outbreak, only the ones that were previously confirmed infected by a laboratory test are included in these datasets.

Moreover, there are relevant variables that are not accurately measured. For example, the fraction of infected non-asymptomatic cases in a given population can be only estimated by means of massive tests or by effective contact-tracing methods. The massive tests carried out in small towns, for example in the north of Italy, indicated that the fraction of asymptomatic cases in the population could be significant (comparable or even larger than the symptomatic cases). Therefore, asymptomatic cases play an important role in virus transmission \cite{YanBai2020}. Furthermore, important inaccuracies have been reported on the use of fast tests. It is an important issue since their use can improve the detection of real cases.

The above limitations on the available datasets have to be taken into consideration in any data-driven method to model or forecast the future spread of the pandemic.

\subsection{Covid-19 mortality}
Being able to predict the number of patients that will develop life-threatening symptoms is important since the disease frequently requires hospitalisation and even ICU in the worst case, challenging the healthcare system capacity \cite{giordano2020modelling}. One of the most important ways to measure the burden of Covid-19 is through mortality. The probability of dying when getting infected depends on different factors \cite{zhou2020clinical}, \cite{peeri2020sars}, \cite{ji2020potential}: 
\begin{itemize} 
\item Demographics \cite{onder2020case}: age, gender, prevalence of diabetes, high blood pressure, obesity, and other risk factors \cite{LEUNG2020111255}. 
\item Health System \cite{ji2020potential}: availability of artificial respiration equipment, ICU, specialized medical surveillance and treatments, etc.
\end{itemize}
\noindent
On the one hand, several studies have reported a higher level of mortality for older people \cite{onder2020case}, even more aggravated in men. Thus, protection strategies should be focused on more vulnerable age and gender groups.

Moreover, the capacity of each regional health system to cope with the pandemic is time-varying. Most of the countries, which had already suffered in a severe way the pandemic, had their hospitals and physicians overwhelmed by the numbers of critical cases (e.g., Italy, Spain, the U.S.) \cite{covid2020forecasting}. The main objective in the control strategies, e.g., contention and mitigation of the disease, is to prevent the saturation or overload of the health system because it will be directly translated into a significant increase in mortality. 

\subsection{Seasonal behaviour of Covid-19}

Many respiratory viruses have a seasonality because lower temperature and lower humidity help facilitate the transmission of the virus \cite{lowen2007influenza} \cite{chan2011effects}. There is no clear evidence that Covid-19 is going to behave seasonally, reducing its transmission in summer. Indeed, during the summer season in the Southern hemisphere, e.g., in some regions of South America and Australia, significant Covid-19 outbreaks have been already reported. In \cite{sajadi2020temperature}, the authors show that on March 2020, the areas with significant community transmission of Covid-19 had distribution roughly along the 30-50º N’ corridor, at consistently similar weather patterns consisting of average temperatures of 5-11ºC and low specific (3-6 g/kg), absolute humidity (4-7 g/m3). In \cite{Wang2020temperature}, the authors study the relationship between temperature, humidity and the transmission rate of Covid-19. They used data collected from all the cities in China with more than 100 cases. The authors use a lineal regression framework as model. Results indicate that increments of one-degree Celsius in temperature and one per cent in relative humidity lower $R_{0}$ by 0.0225 and 0.0158, respectively. The authors developed a web application,\footnote{\url{http://covid19-report.com/\#/r-value}} where $R_{0}$ values for major worldwide cities can be obtained from temperature and humidity.

\subsection{Current actions to control Covid-19 pandemic}

For the control community, the different confinement, pro-active testing and isolation strategies that can be implemented by a government clearly constitute control inputs to the system \cite{anderson2020will}. Many of these strategies to slow or stop the spread of Covid-19 are being implemented worldwide, with different intensities. However, these are not the unique actions that a government can undertake in order to control the pandemic. For example, forcing the population to wear masks (or scarves) and plastic gloves might have an inhibitory effect on the spread of the virus \cite{eikenberry2020mask} and has not a significant impact on the economy (provided masks are produced at large scale). From a control point of view, the objective is twofold. On one hand, it is important to assess the effectiveness of the different measures against the spread of the virus. On the other hand, actions should be planned in advance to mitigate the effects of the pandemic on health system, economy and society. 

It is not simple to determine the effect of the possible anti-measures to be undertaken by the regional governments for several reasons: (i) various inhibitory actions are generally implemented simultaneously, therefore it cannot be evaluated which one has more impact; ii) the efficacy of the anti-measures depends on a number of factors, like demographics and weather conditions of the specific region under consideration; iii) the available data are, in many situations, imprecise and incomplete. The difficulties in predicting the effects of the Covid-19 anti-measures on the regional evolution of disease is one side of the problem. Another one is the inherent time-delay system nature of the dynamics of this disease. The effects of the undertaken measures are observed only weeks later. Another issue is the level of fulfilment of the confinement measures found in each country. In the following, current methods for contention and mitigation of the spread of the virus are described.

\subsubsection{Social distancing}
Following the emergence of this novel coronavirus SARS-CoV-2 and its spread outside China, many countries have implemented unprecedented non-pharmaceutical interventions including case isolation, the closure of schools and universities, banning of mass gatherings and/or public events, and wide-scale social distancing including local and national lockdowns. Many governments around the world closed the educational institutions in an attempt to contain the spread of the Covid-19 pandemic, impacting over 91\% of the world’s student population \cite{UNESCO:Education:Response}. Another important aspect has been tackled by the New York Times: how income affects people’s abilities to stay home and practice social distancing \cite{NYT_social_dist}. Wealthier people not only have more job security and benefits but also may be better able to avoid becoming sick. In \cite{social_distancing}, authors use a semi-mechanistic Bayesian hierarchical model to attempt inferring the impact of these interventions across eleven European countries. They assume that changes in the reproductive number, i.e., a measure of transmission, are an immediate response to these interventions being implemented rather than broader gradual changes in behaviour. In particular, this model estimates these changes by calculating backwards from the deaths observed over time to estimate transmission that occurred several weeks prior, allowing for the time lag between infection and death. One of the key assumptions of the model is that each intervention has the same effect on the reproduction number $R_{0}$ across countries and over time. This allows leveraging a greater amount of data across Europe to estimate these effects. It also means that these results are driven strongly by the data from countries with more advanced epidemics, and earlier interventions, such as Italy and Spain. The main conclusion of this research was that it is critical that the trends in cases and deaths are closely monitored.

\subsubsection{Reducing mobility}
Mobility of people is crucial to understand the spread of the virus. Higher mobility implies higher number of contacts among people \cite{Zhangeabb8001}. Furthermore, national and international mobility explains the rapid spatial propagation of the virus worldwide.  
The authors in \cite{Wang2020mobility} use the Baidu Mobility Index, measured by the total number of outside travels per day divided by the resident population, to find that reducing the number of outings can
effectively decrease the new-onset cases; a 1\% decline in the outing number will reduce about 1\% of the new-onset-cases growth rate in one week (one serial interval). 

Sensor technology can be a crucial tool to obtain mobility measures \cite{oliver2020mobile}. Nowadays, everyone has a mobile phone equipped with a number of sensors, including GPS, that are able to collect data about people mobility. Furthermore, the Internet and mobile phone operators can use their telecommunications towers to gather mobility patterns. Of course, citizen privacy is an issue that has to be taken into consideration for data anonymization. A first quantitative assessment of the impact of the Italian Government on the mobility and the spatial proximity of Italians, through the analysis of a large-scale dataset on de-identified, geo-located smartphone users can be found in \cite{pepe2020covid}.

\subsubsection{Testing} 
The distinction between diagnosed and non-diagnosed is important because non-diagnosed individuals are more likely to spread the infection than diagnosed ones. Indeed, the latter are typically isolated and this can explain mis-perceptions of the case fatality rate and of the seriousness of the epidemic phenomenon \cite{giordano2020modelling}. The main problem for developing massive tests and serology studies is the scarcity of resources, especially in some countries. Accurate testing requires specific labs to analyze RT-PCR tests. On the other hand, the market of rapid tests is under development \cite{kashir2020loop} \cite{nguyen20202019}. Some countries are carrying out serology-based testing. Serology tests are blood-based tests that can be used to identify whether people have been exposed to a particular pathogen by looking at their immune response. In this case, the objective is to have a big picture of the state of population with respect to Covid-19. For instance, to check if herd immunity has been reached in some locations \cite{fine2011herd}.   

\subsubsection{Tracing contacts}
Tracing the contacts of infected people is crucial to isolate potential infected individuals \cite{Ferretti2020}. Once a person is confirmed as an infected one, tracing people contacted with in the last few days can help to reduce the propagation of virus. However, tracing contacts results to be a challenging task. Manual registers can require an amount of resources unaffordable for the majority of countries. Therefore, technology should play an important role \cite{rao2020identification} \cite{Ferretti2020}, in particular mobile devices \cite{oliver2020mobile2} and wireless technologies, such as WiFi and Bluetooth.  

%% file: Data_driven_ways_to_fight_the_pandemic.tex
\section{Data driven techniques to fight the pandemic}\label{sec:Data:Driven}
Currently, the majority of data available on Covid-19 is used for describing the pandemic in terms of reports and visualizations.\footnote{For example, \url{https://againstcovid19.com/singapore/dashboard}} Although these techniques are useful to highlight the magnitude of the crisis, they are not enough for contending and mitigating the problem. Also, these are insufficient for decision-makers to anticipate the response to the virus propagation and evaluate the effectiveness of the implemented actions. Classic epidemic models are also useful to obtain mathematical models for epidemics \cite{Martcheva2015}. However, many parameters of these models, such as infected rate and basic reproduction number, require data-driven approaches to estimate them accurately. Also, classic epidemic models, which are normally based on curve fitting techniques, require data on different phases of the epidemic to obtain the parameters. For these reasons, it is obvious that more efficient approaches are needed rapidly to: i) model and forecast the spread and the consequences of the pandemic; and ii) evaluate mitigation approaches that have been carried out. Data-driven models (see ,e.g., \cite{giordano2020modelling} and \cite{Gomez2020modelado}) can be such solution \cite{zhang2020estimation}, \cite{fang2020transmission}. Many data-based techniques can be applied \cite{mahalle2020forecasting}, ranging from classical statistical and machine learning approaches, e.g., linear regression \cite{perone2020arima} \cite{calafiore2020modified} and Bayesian inference \cite{flaxman2020estimating}, to sophisticated models based on neural networks \cite{Dandekar2020neural}. These techniques require sufficient and high-quality data to provide a good estimation. Depending on the methodology used, the quantity of data can vary notably from hundreds to millions of samples. Moreover, a wide variety of data can be necessary for accomplishing an accurate model of a complex and dynamic system like the Covid-19 pandemic. Therefore, data from different disciplines are required, which hinders the data collection task. We highlight three pillars of data-driven approaches for fighting Covid-19: i) informative variables for developing an accurate model; ii) objectives of the model: characterizing Covid-19 pandemic, epidemic models and forecasting, etc.; and iii) its use for efficient decision making. 

\begin{itemize}
    \item \textbf{Wish-list of variables:} the list of variables is large, since many aspects should be taken into consideration to develop accurate models. The considered variables can be divided into different categories, according to their discipline.\footnote{The following list can be improved including other disciplines and variables.}
    \begin{itemize}
        \item \textit{Covid-19 variables}: regional time series of the number of confirmed cases, suspicious cases, deaths, recovered, number of tests, hospitalized cases, ICU cases, isolated positive cases, serology studies, etc. When possible, the data should be divided per gender, age range, etc. \\
        \item \textit{Geographic variables}: locations of Covid-19 variables. The locations can be obtained from either names, e.g., countries, cities, etc., or GPS coordinates, i.e., longitude and latitude.\\
        \item \textit{Demographic variables}: population and density of population by location. These variables are required for normalization of the rest of the variables. Other parameters are the age structure of the population, the prevalence of secondary health conditions related to higher Covid-19 mortality, etc. \\
        \item \textit{Health system variables}: total number of ICU beds, number of doctors and nurses, personal protective equipment (PPE), respirators, number and types of tests. \\
        \item \textit{Government measures}: social distancing, movement restrictions, lockdowns, etc. \\
        \item \textit{Weather variables}: temperature, relative humidity, radiation, etc.\\
        \item\textit{ Contamination variables}: air pollution, i.e., fine particulate matter $PM_{2.5}$.\\
        \item \textit{International and national mobility and connectivity}: number of international and national flights, number of train connections international and national mobility patterns, traffic patterns, etc. 
    \end{itemize}
    \item \textbf{The use of data to estimate the state of the epidemic and develop forecasting models:} By using the aforementioned variables, different models can be developed to estimate the current state of the pandemic and anticipate the response to the propagation of Covid-19. Examples of estimation and forecasting analyses are:
    \begin{itemize}
        \item Estimation of the infected population.\\
        \item Estimation of economic impact.\\
        \item Forecast of impact in health system through number of infected.\\
        \item Assessing the impact in terms of mortality.\\
        \item Analysis of seasonal behaviour.\\
    \end{itemize}
    \item \textbf{Decision making:} The final objective of the data-driven models is developing useful tools for helping governments and institutions to anticipate the response to the Covid-19 propagation and evaluate their actions. Among them, the most relevant are:
    \begin{itemize}
        \item Assessing the effectiveness of the measures.\\
        \item Planing ahead government actions. 
    \end{itemize}
\end{itemize}

%% file: Limitations_and_challenges_raised_by_the_available_data.tex
\section{Limitations and challenges raised by the available data}\label{Sec:Limitations}
There exist different issues that can hinder the use of open data to address the challenges raised by the Covid-19 pandemic. The main obstacles are addressed in the following sections.

\subsection{Variety of formats} Since there is no a common shared open database on Covid-19, the different sources and variables required to undertake a given analysis are often addressed by assembling several data sets into a single one. Although the increased quantity of data sources presents new opportunities, working with such a variety of data reinforces the validity challenges \cite{mooney2015epidemiology}. Another issue is related to the wide range of disciplines from which the data sources are coming from. Indeed, these disciplines can be familiar with very different formats and data representation. For instance, some available APIs (Application Programming Interfaces) to get data on Covid-19 provide JavaScript Object Notation (JSON) files. This format is widely used in computer science for web applications. However, for instance, mathematicians and epidemiologists could not be familiar with such format. 

\subsection{Time-varying nature}
The needs of the outbreak require immediate response, which translates in obtaining the latest information available. This raises some important challenges. For example, government measures are changing rapidly. Often information is outdated by the time it has been identified. The number of countries implementing or amending measures increases daily \cite{Acaps:GovernmentMeasures:Report}. 
The daily availability of the data can be an issue for working with multiple data sources simultaneously. 

\subsection{Confirmed cases is not a reliable metric}
In the WHO global \href{https://www.who.int/publications-detail/global-surveillance-for-human-infection-with-novel-coronavirus-(2019-ncov)}{Covid-19 surveillance document}, a confirmed case is defined as a person with laboratory confirmation of Covid-19 infection, irrespective of clinical signs and symptoms. At the outbreak of the pandemic, the access to massive tests was very limited and often only a reduced fraction of the hospitalized cases were tested at a laboratory level. Thus, most reports of infection are extremely filtered by the complex and limited testing process. Furthermore, very few datasets provide information about the number of suspected cases.

Even under the hypothesis that everyone with minor symptoms is tested, this would only provide an estimate of the symptomatic cases of the disease. The study of the fraction of asymptomatic cases is an active field of research (see e.g., \cite{mizumoto2020estimating} and \cite{Ferretti2020}) not only because it is one key to the estimation of the total number of infected cases, but because it plays a fundamental role in the spread of the virus \cite{Ferretti2020}. 

\subsection{Mortality rate is difficult to estimate }
During the most severe periods of the virus spread in a country, in many situations the number of death cases reported by the administration differs considerably from the real one. This is because only the deaths with previous laboratory confirmation of the disease are included. Thus, the study of national death registers suggests that there are notably and unexpected increases in death rates, according to the historical numbers. For instance, New York City has reported 5330 more deaths than expected in April 2020\footnote{\url{https://www.nytimes.com/interactive/2020/04/10/upshot/coronavirus-deaths-new-york-city.html}}, only 3350 of these can be accounted for Covid-19 reasons. These figures suggest that there exist an undercounting on the real number of deaths. Another example is reported for Spain, where the \textit{"Sistema de Monitorización de la Mortalidad diaria (MoMo)}\footnote{\url{https://www.isciii.es/QueHacemos/Servicios/VigilanciaSaludPublicaRENAVE/EnfermedadesTransmisibles/MoMo/Paginas/MoMo.aspx}} system registers the total number of deaths under any circumstance. The report on April 7th indicates an increase of more than 50\% of unexpected deaths in the month before. Such increment is even more significant in men, where it reaches more than 60\%.

The mortality rates are much more difficult to estimate since the estimates are often based on the number of deaths relative to the number of confirmed cases of infection, which can be a small fraction of the real ones \cite{baud2020real}. Consequently, the comparison of mortality rates between countries makes compulsory the implementation of correcting factors based on the estimation of Covid-19 infected cases and deaths non registered by the respective administrations. Also, when considering the increase of mortality due to saturation of the health care system, one has to take into consideration the fact that the patients who die on any given day were infected much earlier. Thus, the denominator of the mortality rate should be the total number of patients infected at the same time as those who died \cite{baud2020real}. 

Another important parameter to evaluate mortality is to have stratified data, according to gender and age groups. However, such information is not provided by the majority of data sources. 

\subsection{Not availability of individual case data}
To better understand the disease and to improve models and strategies to fight Covid-19, each case should be tracked with its own timeline. That is, for each case, relevant information about when symptoms appeared, medical treatments, evolution, degree of isolation, etc., should be available on a country-wide level. Then, this data should be published anonymously, with a de-identification process, to prevent personal identity from being revealed. The data, and the time corresponding to the change of each individual, should be published by an official source in a structured way, at least, with daily frequency. This possibility is supported by the opinion of many experts and members of the open-source community.\footnote{See, for example, \url{https://github.com/jgehrcke/covid-19-germany-gae}.} 

An effort of obtaining individual case data can be found in \cite{oliver2020covid19impact}. The authors carried out a survey of 24 questions related to the impact of Covid-19 (Covid19Impact) on citizens in Spain. \footnote{\url{https://survey123.arcgis.com/share/d29378b51fe8496d8dd77f08ce73973f}} The survey was responded to by 146,728 participants over a period of less than two days (i.e., 44 hours). The questions were about social contact behaviour, financial impact, working situation, and health status. The results of the survey show the negative impact of Covid-19 on the life of citizens. It is a clear example of how the collaboration of the citizens can be relevant to gather information on the effects of Covid-19. A similar work has been pursued in UK and the results can be found in \cite{Kleinberg2020measuring}, where the authors created the Real World Worry Dataset of 5,000
texts.\footnote{\url{https://github.com/ben-aaron188/covid19worry}} The data analysis suggests that people in the UK specially worry about their family and the economic situation.  

\subsection{Changing and non-uniform criteria}

Since the governments are continuously adjusting their response to the virus, it is common to find out abrupt changes in the trend of a timeseries because a new methodology has been implemented. For example, on February 12th, a sudden spike of 15,152 new Covid‐19 cases in China was observed and it was related to the modified method used for diagnosis, i.e., a combination of SARS‐CoV‐2 nucleic acid test and clinical Covid‐19 features \cite{wang2020combination}.

Another relevant issue is that regions in the same country may provide data under the same label, but with a different meaning. A good example is represented by the number of ICU cases. There might be regions reporting the accumulative number of confirmed cases that required ICUs, and others the number of ICUs used by Covid-19 patients. Something similar happens with the number of laboratory tests. They can refer either to the total number of tests carried out or to the number of individuals tested. Indeed, in many situations, the sources do not describe accurately the meaning of the counts.

\subsection{Changing data-base structure and locations}

The open data sources on Covid-19 are constantly improving. To provide more meaningful information, new variables are incorporated into datasets. This translates into a change in the structure of the data, which requires adjusting the code to download and process the information. When regional data are collected from the official open data portals of different countries, a surveillance effort is required to keep track of the different modifications. In many situations, the new data-files appear in different locations with different names. 

\subsection{Government transparency}
There are important differences in how the governments are reporting the data related to Covid-19. Furthermore, there are some concerns about the transparency of countries regarding the data provided.

\subsection{Rush in academia publications}
Many scientific papers are being rapidly published even without peer-review, which is a sub-optimal way to publish science, and more studies are being based on data that is essentially non-peer-reviewed that may have a potential for bias or may contain genuine errors in research methodologies.

%% file: Open_data_institutions_providing_worldwide_Covid_19_data.tex
\section{Open data institutions providing worldwide Covid-19 data }\label{sec:Global:OpenDataInstitutions}

Numerous institutions of different nature, e.g., global institutions, European Union (EU) institutions, universities, newspapers, etc., are providing daily reports on the evolution of the Covid-19 pandemic. In this section, we enumerate those that, from our experience, resulted to be the most relevant and reliable ones. In particular, we highlight the ones that provide updated information on a regular basis in the open-data repository with easy access. Some of the enumerated institutions are making a great effort to provide consolidated data, describing in a rather exhaustive form, the sources and limitations of the provided datasets. In this section, we describe the nature and characteristics of the information provided, detailing the specifics of the datasets only for the most relevant ones.

\subsection{World Health Organization}
The primary role of WHO is to direct international health within the United Nations' system and to lead partners in global health responses. In the framework of the pandemic Covid-19, WHO is providing continuous updates about the current situation all around the world.\footnote{\url{https://www.who.int/westernpacific/emergencies/covid-19}} In \cite{WHO_Situation_report}, WHO provides guidelines to follow, in the privacy of our house as well as in public, Q\&A pages on the most common questions about the virus, how it spreads and how it is affecting people worldwide,. Moreover, it addresses also myth busters related to Covid-19, in order to provide a reliable source of information (see \cite{WHO_myth}).

\subsection{Johns Hopkins University}
Johns Hopkins experts in global public health, infectious disease, and emergency preparedness have been at the forefront of the international response to Covid-19 (see \url{https://coronavirus.jhu.edu/}) since the beginning. This university provides a \href{ https://coronavirus.jhu.edu/map.html}{daily update} on the global map of the pandemic. The dataset provided by the Johns Hopkins University (JHU) (see sub-subsection \ref{SubSubSection:JHU:Data:Set}) is one of the most frequently used by researchers and journal media. 

\subsection{University of Oxford}
The Blavatnik School of Government is a department of University of Oxford that is working on the Covid-19 pandemic and on the policy responses we see around the world. One of their projects related to the study of Covid-19 is focused on tracking what governments around the world are responding to the pandemic and how they compare to others.\footnote{Further information on the actions developed can be found at:
\url{https://www.bsg.ox.ac.uk/news/coronavirus-research-blavatnik-school}.} Regarding the comparison of confinement strategies developed by governments, they have created a common index named \textit{Stringency Index}. This index is based on data obtained by the Oxford Covid-19 Government Response Tracker (OxCGRT), which systematically collects information on several different common policy responses governments have taken.

\subsection{European Union}
The European Data Portal (EDP)\footnote{\url{https://data.europa.eu/euodp/es/data/}}, which is the official open data portal of the European Union, gives access to open data published by EU institutions and bodies. EDP acts as single access point to open data and it is published by national open data portals and institutions in the EU Member States as well as by other non-EU countries. There are numerous datasets on EDP that reference "covid" or "corona". Also, less specific datasets describing former health infections, epidemics or pandemics are also provided.\footnote{\url{https://www.europeandataportal.eu/en/highlights/covid-19}}

In an effort to promote research on Covid-19, the European Union has opened a specific data portal, called COVID-19 Data Portal \url{https://www.covid19dataportal.org/}. The datasets included in the portal are divided into six categories, such as sequences, expression data, protein, structures, literature and other resources. 

In the follows, some of the most relevant European research centers, which have been tackling with the Covid-19 outbreak, are briefly presented.

\subsubsection{Joint Research Centre}
The Joint Research Centre (JRC) is the European Commission's science and knowledge service,\footnote{\url{https://ec.europa.eu/knowledge4policy/organisation/jrc-joint-research-centre_en}} which employs scientists to carry out research in order to provide independent scientific advice and support to EU policy. 

\subsubsection{European Center for Disease Prevention and Control}
The European Center for Disease Prevention and Control (ECDC), established in 2004 after the 2003 SARS outbreak and located in Solna, Sweden, is an independent EU agency, whose mission is to strengthen Europe's defences against infectious diseases. ECDC publishes numerous scientific and technical reports covering various issues related to the prevention and control of infectious diseases. Towards the end of every calendar year, ECDC publishes its Annual Epidemiological Report, which analyses surveillance data and infectious disease threats. In addition to offering an overview of the public health situation in the EU, the report offers an indication of where further public health action may be required to reduce the burden caused by communicable diseases. As other organizations, ECDC is closely monitoring the Covid-19 pandemic, providing risk assessments, public health guidance, advice on response activities to EU Member States and the EU Commission, and daily-updated data on current outbreak \cite{ecdc_data}.

For EU level surveillance, ECDC requests countries from EU and from the European Economic Area (EEA) and UK to report laboratory-confirmed cases of Covid-19 within 24 hours after identification. This is done through the Early Warning and Response System (EWRS).

\subsubsection{European Centre for Medium-Range Weather Forecasts}
The European Centre for Medium-Range Weather Forecasts (ECMWF) is an independent intergovernmental organization supported by 34 states based in Reading \cite{ECMWF}. ECMWF is both a research institute and a 24/7 operational service, producing and disseminating numerical weather predictions to EU Member States, Co-operating States and the broader community. ECMWF also archives data and makes them available to authorized users. Some data are also made available under licence, and some are publicly available. 

\subsection{United Nations}
Good examples of open-data provided by the United Nations (UN) are reported in \cite{UN_web}. Moreover, \cite{CSSE} contains the most up-to-date Covid-19 cases and latest trend plot. It covers China, Canada, Australia at province/state level whereas the rest of the world, including US, is covered at country level, represented by either the country centroids or their capitals.

\subsection{The New York Times}
The New York Times is releasing a series of data files with cumulative counts of Covid-19 cases in the U.S., at state and county level, over time. The timeseries data are compiled from states, local governments and health departments. Since January 2020, The NY Times has tracked cases of coronavirus in real-time as they were identified after testing. Then, these data have been used to power maps and generate reports about the outbreak. The data collection began with the first reported coronavirus case in Washington State, on January 21st, 2020. Since then, the NY Times publishes regular updates of data in a \href{https://github.com/nytimes/covid-19-data}{GitHub  repository}.

\subsection{Our World In Data}
Our World in Data (OWID) is an online scientific publication that focuses on large global problems, such as poverty, disease, hunger, climate change, war, existential risks, and inequality. Covid-19 data provided by OWID can be found at their \href{https://ourworldindata.org/coronavirus}{open-data portal}. 

\subsection{Africa Centres for Disease Control and Prevention}
Africa Centres for Disease Control and Prevention (CDC) is a specialized technical institution of the African Union established to support public health initiatives of Member States and strengthen the capacity of their public health institutions to detect, prevent, control and respond quickly and effectively to disease threats.\footnote{\url{https://africacdc.org/}} They provide reports on status, mitigation strategies and guidelines on Covid-19 at \url{https://africacdc.org/covid-19/covid-19-resources/}. 

\subsection{Google}
The multinational technology company Google has developed a visual Covid-19 map, where also relevant information can be found, worldwide and by country \url{https://google.com/covid19-map/}. The map is continuously updated and the data exploited are taken from Wikipedia.\footnote{\url{https://en.wikipedia.org/wiki/Template:2019\%E2\%80\%9320_coronavirus_pandemic_data}} They also present statistics about the number of confirmed cases, cases per one-million of people (normalized data), number of people recovered, and deaths. 

Another relevant tool developed by Google, which can be used to obtain data about Covid-19, is the Google DataSet Search.\footnote{ \url{https://datasetsearch.research.google.com/}} Numerous data sets can be found looking for the term \textit{Covid-19}. The application allows users to filter the datasets by several fields, such as last updated, download format, usage rights, topic, and accessibility, etc. 

\subsection{ACAPS}\label{sub:sec:ACAPS} 
ACAPS, initially known as The Assessment Capacities Project, is an independent information provider helping humanitarian actors respond more effectively to disasters (\url{https://www.acaps.org}). ACAPS was established in 2009 as a non-profit, non-governmental project with the aim of providing independent, ground-breaking humanitarian analysis to help humanitarian workers, influencers, fundraisers, and donors make better decisions. It is not affiliated to the UN or any other organization but is a non-profit project of a consortium of two NGOs, i.e., the Norwegian Refugee Council and Save the Children, and it receives support from several international sources, e.g., the Humanitarian Aid and Civil Protection organization. The ACAPS analysis team is mainly dedicated to researching and analyzing global and crisis specific data. 
They provide regional reports on the pandemic, and additional information like description of the worldwide measures against the spread of the virus available at \url{https://www.acaps.org/what-we-do/reports} and in \cite{Acaps:GovernmentMeasures:Report}. 

\subsection{ Organization for Economic Co-operation and Development}
The Organisation for Economic Co-operation and Development (OECD)\footnote{\url{https://www.oecd.org}} is an international organization that, together with governments, policymakers and citizens, has the goal of establishing evidence-based international standards and finding solutions to a range of social, economic and environmental challenges. From improving economic performance and creating jobs to fostering strong education and fighting international tax evasion, they provide a forum and knowledge-hub for data and analysis, experiences exchange, best-practice sharing, and advice on public policies and international standard-setting. OECD provides different reports and data about government actions and economic impact due to the pandemic, which can be found at \url{http://www.oecd.org/coronavirus/en/}. 

\subsection{Medical Research Council Centre for Global Infectious Disease Analysis}
The Medical Research Council Centre for Global Infectious Disease Analysis (MRC GIDA) of the Imperial College of London is an international resource and centre of excellence for research and capacity building on the epidemiological analysis and modelling of infectious diseases, and to undertake applied collaborative work with national and international agencies to support policy planning and response operations against infectious disease threats. The MRC presents reports on Covid-19 under five categories:\footnote{\url{https://www.imperial.ac.uk/mrc-global-infectious-disease-analysis/covid-19/}} i) weekly-forecasts; ii) resources; iii) information; iv) video updates; and v) publications.

Furthermore, in collaboration with several departments of Imperial College London (Imperial College Covid-19 Response Team) and Oxford University, they developed a model\footnote{The updates of the model can be accessed at \url{https://github.com/ImperialCollegeLondon/covid19model}.} for estimating the number of infections and the impact of non-pharmaceutical interventions on Covid-19 in eleven European countries \cite{flaxman2020estimating}. 

\subsection{The Institute for Health Metrics and Evaluation}
The Institute for Health Metrics and Evaluation (IHME) is an independent global health research center at the University of Washington.\footnote{\url{http://www.healthdata.org/}} They have developed a model to determine the extent and timing of deaths and excess demand for hospital services due to Covid-19 in the US \cite{covid2020forecasting}. The work uses: i) data on confirmed Covid-19 deaths from WHO and from local and national governments; ii) data on hospital capacity and utilization for US states; and iii) observed Covid-19 utilization data from different locations. A web service, where the projections of the model can be determined for each country and for the following four months, is available.\footnote{\url{https://covid19.healthdata.org/projections}} The information provided is: i) hospital resources needs, including the number of beds, the number of ICU beds, and ventilators; ii) the number of death per day; and iii) the total number of deaths.

\subsection{ New England Complex Systems Institute (NECSI)}
It is a research institution in USA. Its main focus is on advancing the study of complex systems.\footnote{\url{https://necsi.edu/}} They have developed a \href{https://www.endcoronavirus.org/}{portal} with the following goals: i) stop the spread of COVID-19, ii) consult governments, iii) institutions and individuals, iv) provide useful data and guidelines, and v) crush the curve. The portal includes guidelines and reports on governments, communities, medical institutions, companies, families and individuals.

\subsection{MIDAS Network}
MIDAS is a global network of scientists and practitioners from academia, industry, government, and non-governmental agencies, who develop and use computational, statistical and mathematical models to improve the understanding of infectious disease dynamics as it relates to pathogenesis, transmission, effective control strategies, and forecasting.\footnote{\url{https://midasnetwork.us/covid-19/}} They have created a portal for Covid-19 modeling, which provides an important and reliable catalogue of data resources, including datasets, webinars, and funding announcements.

\subsection{COVID-19 Data Hub}
The COVID-19 Data Hub projct has been funded by the Institute for Data Valorization IVADO, Canada.\footnote{\url{https://ivado.ca/en/}} The goal of the project is to provide the research community with a unified data hub by collecting worldwide fine-grained case data merged with demographics, air pollution, and other exogenous variables helpful for a better understanding of COVID-19.\footnote{\url{https://covid19datahub.io/}} In addition, they provide R package to download Covid-19 related datasets.

\subsection{Science.gov}
Science.gov, a gateway portal to U.S. government science information with free access to research and development results and scientific and technical information from scientific organizations across 13 federal agencies, uses software that supports federated search in real-time, over 70 information sources (e.g., databases) across the leading federal science and technology agencies in the United States. Using a combination of search terms for Covid-19, Science.gov has provided a \href{https://www.science.gov/coronavirus.html}{link} off its homepage that the public can use to quickly access federally-funded research on the COVID-19 disease. Upon linking to the coronavirus research results, users can access freely available peer-reviewed literature (journal articles and accepted manuscripts).

\subsection{United States National Institute of Standards and Technology}
The National Institute of Standards and Technology (NIST) 
is a physical sciences laboratory and a non-regulatory agency of the United States Department of Commerce. Its mission is to promote innovation and industrial competitiveness. NIST's activities are organized into laboratory programs that include also information technology. For the Covid-19 pandemic, they provide a dedicated open portal where it is possible to search for specific \href{randr19.nist.gov}{datasets} related to the virus outbreak. Moreover, in collaboration with Allen Institute for Artificial Intelligence (AI2), the National Library of Medicine (NLM), Oregon Health \& Science University (OHSU), and the University of Texas Health Science Center at Houston (UTHealth), NIST has formed the so-called TREC-COVID challenge, which is currently building a set of Information Retrieval (IR) test collections based on the CORD-19 datasets (see Section 6.3 for further details on CORD-19 competition) and the Text Retrieval Conference (TREC) model. Additional information on this challenge can be found at \url{ https://ir.nist.gov/covidSubmit/}.

\subsection{United States National Institutes of Health}
The National Institutes of Health, which represents the primary agency of the United States government responsible for biomedical and public health research, is one of the most prominent source of data on Covid-19 pandemic\footnote{\url{https://www.nih.gov/health-information/coronavirus}}. In particular, the NIH Office of Data Science Strategy provided a portal dedicated to open-access data and computational resources related to Covid-19 fight available at \url{https://datascience.nih.gov/covid-19-open-access-resources}, seeking to provide the research community with links to open-access data (see e.g., \url{https://www.ncbi.nlm.nih.gov/pmc/about/covid-19/}), computational, and supporting resources. 

\subsection{Open Data Watch}
Open Data Watch is a non-profit, non-governmental organization founded by three development data specialists.\footnote{\url{https://opendatawatch.com/}} It monitors progress and provides information and assistance to guide the implementation of open data systems. The Open Data Watch team is experienced in the development of data management and statistical capacity-building in developing countries. They have collected data from different sources all around the world related to the Covid-19 pandemic. Indeed, to address the ongoing need for data-driven decision making, Open Data Watch has put together some articles, organized by the stages of the data value chain: availability, openness, dissemination, and use and uptake. These papers are updated as new information becomes available. These references and related links can be found in \cite{OpenDataWatch}.

\subsection{EuroMOMO}
It is a European mortality monitoring activity, aiming to detect and measure excess deaths related to seasonal influenza, pandemics and other public health threats\footnote{\url{https://www.euromomo.eu/}}. They report weekly \href{https://www.euromomo.eu/bulletins/2020-18/}{bulletins} on excess of mortality of European countries.

\subsection{World Bank Open Data}
The World Bank Group (WBG) is a family of five international organizations that make leveraged loans to developing countries. The World Bank's activities are mainly focused on developing countries, in fields such as education, health, agriculture, etc. During the Covid-19 pandemic, WBG help developing countries strengthen their pandemic response and health care systems. Furthermore, WBG has highlighted the importance of \textit{data} to support countries in managing the global Covid-19 outbreak, including in their open data portal, i.e., the World Bank Open Data,\footnote{\url{https://data.worldbank.org/}} an entire \href{https://www.worldbank.org/en/who-we-are/news/coronavirus-covid19?intcid=wbw_xpl_banner_en_ext_Covid19}{section} dedicated to Covid-19 and \href{http://datatopics.worldbank.org/universal-health-coverage/coronavirus/}{datasets} with real-time data, statistical indicators, and other types of data that are relevant to the coronavirus pandemic, particularly focused on the economic and social impacts of the pandemic and the World Bank’s efforts to address them.

This dataset is of particular relevance to assess the correlation among the health emergency and the extraordinary shock the global economy is facing, trying to reply to the question: \textit{how is the deadly virus impacting global poverty?}.\footnote{\url{https://blogs.worldbank.org/opendata/impact-covid-19-coronavirus-global-poverty-why-sub-saharan-africa-might-be-region-hardest}} Indeed, estimating how much global poverty will increase because of COVID-19 is challenging and comes with a lot of uncertainty. To answer this question, they propose a model based on household survey data provided by \href{http://iresearch.worldbank.org/PovcalNet/povOnDemand.aspx}{PovcalNet} (an online tool provided by the World Bank for estimating global poverty) and extrapolate forward using the growth projections from the recently launched World Economic Outlook. Comparing these Covid-19-impacted forecasts with the forecasts from the previous edition of the World Economic Outlook provides an assessment of the impact of the pandemic on global poverty, assuming that the pandemic does not change inequality within countries.

%% file: Open_source_communities.tex
\section{Open source communities }\label{sec:Open:Data:Repositories}

This section covers repositories of open source communities, which are dedicated to joining people with similar interests. These have been widely developed in the software field, where many professionals and practitioners join their efforts to achieve bigger goals on software projects. These communities are playing a very active role in facilitating access to Covid-19 datasets from official open portals all over the world. 

\subsection{GitHub}
GitHub is a subsidiary company of Microsoft for hosting software development using Git. It provides control versions and project management, among other tools. Numerous open software projects are daily posted, free of charge. Since Covid-19 outbreak, many projects and related datasets have been posted. The majority of those included in this paper can be obtained from GitHub. Some examples are: i) \href{https://github.com/open-covid-19/data}{Open Covid-19 Dataset}; ii) \href{https://github.com/covid19-data/covid19-data}{Covid-19 Data Processing Pipelines and Datasets}; and iii) \href{https://github.com/pomber/covid19}{JSON timeseries of coronavirus cases dataset}.

\subsection{Harvard Dataverse Repository}
Harvard Dataverse is a free data repository, open to all researchers from any discipline, both inside and outside of the Harvard community. Researches and practitioner can share, archive, cite, access, and explore research data.\footnote{\url{https://support.dataverse.harvard.edu/}} They have opened a link at \url{https://dataverse.harvard.edu/dataverse/2019ncov} for works related to Covid-19, where both the papers and the data used for the analysis can be found.

\subsection{Kaggle}
Kaggle is a community for data scientist and machine learning practitioners. Kaggle allows users to find and publish datasets, to explore and build models in a web-based data-science environment, to work with other data scientists and machine learning engineers, and to enter competitions to solve data-science challenges. Regarding Covid-19 pandemic, the portal opens a new challenge weekly to work on Covid-19 data.\footnote{\url{https://www.kaggle.com/tags/covid19}} The challenge consists of forecasting confirmed cases and fatalities for the following week. Furthermore, some data analysis posts can be found for each competition.\footnote{For example  \url{https://www.kaggle.com/frlemarchand/covid-19-forecasting-with-an-rnn}.} The challenges opened up to the date (10/04/2020) can be found at the following links:
i) March 18th: \url{https://www.kaggle.com/c/covid19-global-forecasting-week-1};
ii) March 25th: \url{https://www.kaggle.com/c/covid19-global-forecasting-week-2};
iii) April 1st: \url{https://www.kaggle.com/c/covid19-global-forecasting-week-3};
and iv) April 8th: \url{https://www.kaggle.com/c/covid19-global-forecasting-week-4}. 

Moreover, the Covid-19 Open Research Dataset Challenge (CORD-19) competition has been launched, aimed at developing text and data mining tools that can help the medical community to develop answers to high priority scientific questions \url{https://www.kaggle.com/allen-institute-for-ai/CORD-19-research-challenge}. The available dataset, based on data sources provided by the \href{http://cset.georgetown.edu}{Center for Security and Emerging technology of Georgetown University} is composed by a corpus of more than 44,000 full-text documents, about Covid-19/SARS-CoV-2 and related coronaviruses. 

Another relevant competition based on Covid-19 data is the UNCOVER COVID-19 Challenge \url{https://www.kaggle.com/roche-data-science-coalition/uncover}. In this case, the objective is modeling solutions to key questions that were developed and evaluated by a global front-line of healthcare providers, hospitals, suppliers, and policy makers. In this case, the challenge is promoted by Hoffmann-La Roche Limited (Roche Canada).

\subsection{Zindi}
Zindi is the first data-science competition platform in Africa. Zindi hosts an entire data-science ecosystem of scientists, engineers, academics, companies, NGOs, governments and institutions, focused on solving Africa’s most pressing problems. Regarding Covid-19 pandemic, they have open a competition aimed at building an epidemiological model that predicts the spread of Covid-19 throughout the world. The target variable is the cumulative number of deaths caused by COVID-19 in each country by each date. The challenge can be found at \url{https://zindi.africa/competitions/predict-the-global-spread-of-covid-19/data}.

%% file: Covid_19_Data_Sets.tex
\section{Covid-19 Datasets}\label{sec:Covid19:DataSets}
This section presents the main available datasets that can be found on the Internet related to Covid-19. The section is divided into two parts. First, we present international datasets that provide global information related to the virus impact of each country, such as number of total/new confirmed cases and number of total/new confirmed death. Second, we include a number of regional data sets, where local information can be found. Although the information can be redundant on several data sets, we believe that it could be interesting to validate the developed models/analysis.   

\subsection{International datasets}
In this section, we briefly introduce the institutions that provide international datasets, including also the link (URL) to an easier access to them.

\subsubsection{Johns Hopkins University Data Set}\label{SubSubSection:JHU:Data:Set}
Johns Hopkins experts in global public health, infectious disease, and emergency preparedness have been at the forefront of the international response to Covid-19.\footnote{\url{https://coronavirus.jhu.edu/}} JHU provides a daily update of the global map of the pandemic, which can be found at \url{https://coronavirus.jhu.edu/map.html}.
    
The JHU Covid-19 dataset can be downloaded in .csv format from the Github \href{https://github.com/CSSEGISandData/COVID-19/tree/master/csse_covid_19_data/csse_covid_19_time_series}{repository}. In this folder, five different .csv files can be downloaded: i) \href{https://github.com/CSSEGISandData/COVID-19/blob/master/csse_covid_19_data/csse_covid_19_time_series/time_series_covid19_confirmed_global.csv}{global number of confirmed cases}; ii) \href{https://github.com/CSSEGISandData/COVID-19/blob/master/csse_covid_19_data/csse_covid_19_time_series/time_series_covid19_deaths_global.csv}{global number of deaths}; iii) \href{https://github.com/CSSEGISandData/COVID-19/blob/master/csse_covid_19_data/csse_covid_19_time_series/time_series_covid19_recovered_global.csv}{global number of recovered}; iv) \href{https://github.com/CSSEGISandData/COVID-19/blob/master/csse_covid_19_data/csse_covid_19_time_series/time_series_covid19_confirmed_US.csv}{total number of confirmed cases in US}; and v) \href{https://github.com/CSSEGISandData/COVID-19/blob/master/csse_covid_19_data/csse_covid_19_time_series/time_series_covid19_deaths_US.csv}{total number of deaths in US}. The global files refer to worldwide Covid-19 data. A reduced number of countries are further divided into regions, e.g., China and Australia, whereas, most of them like Spain or Italy, are not. The U.S. data .csv files correspond to the United States. In both cases, all the data refer to accumulated cases, i.e., cases up to the date of the row in which the data is consigned. Furthermore, the geographical coordinates of each region/country are also provided. 

The data plots, which can be recovered at \url{https://coronavirus.jhu.edu/data/new-cases}, are obtained by means of a 5-days moving window, averaging the values of that day, the two days before, and the two days after. This approach helps to avoid major events, such as a change in reporting methods, from skewing the data. 

\subsubsection{ Geographical Distribution of Covid-19 Worldwide}
The Geographical Distribution of Covid-19 Worldwide Dataset is sourced from the ECDC, which publishes full timeseries data for the number of confirmed Covid-19 cases and deaths daily for various countries around the world. On daily basis, the ECDC collects data from 6am to 10am CET and publishes this data via its Covid-19 dashboard.\footnote{\url{https://qap.ecdc.europa.eu/public/extensions/COVID-19/COVID-19.html }} Then, this dataset is also made publicly available through downloadable files in different formats.\footnote{\url{https://www.ecdc.europa.eu/en/publications-data/download-todays-data-geographic-distribution-covid-19-cases-worldwide}} This dataset can be also downloaded from the open-portal of \textit{Our World in Data}. In particular, it is possible to recover the following datasets: i) \href{https://covid.ourworldindata.org/data/ecdc/total_cases.csv}{total confirmed cases}; ii) \href{https://covid.ourworldindata.org/data/ecdc/total_deaths.csv}{total deaths}; iii) \href{https://covid.ourworldindata.org/data/ecdc/new_cases.csv}{new confirmed cases}; iv) \href{https://covid.ourworldindata.org/data/ecdc/new_deaths.csv}{new deaths}; v) \href{https://covid.ourworldindata.org/data/ecdc/full_data.csv}{all four metrics}; and vi) \href{https://covid.ourworldindata.org/data/ecdc/locations.csv}{population data}.

\subsubsection{Covid-19 Data Hub}
The Covid-19 Data Hub project makes all the data available at \url{https://github.com/covid19datahub/COVID19}. The dataset includes a large ranges of variables such as Covid-19 variables (confirmed cases, death, etc.), population, density, ICU, number of tests, ventilators, testing policy and contact tracing, among others. 

\subsubsection{MIDAS Network Dataset}
The MIDAS network publishes an open dataset with several data resources to study Covid-19 pandemic.\footnote{\url{https://github.com/midas-network/COVID-19}} The resources are divided into different sections, such as data catalog, parameter estimates, software tools, and documents. In particular, a collection of .csv files can be found in the catalog section about the situation of each country.

\subsubsection{Covid-19 Testing ({\it Our World in Data} Dataset)}
{\it Our World in Data} publishes useful information on how the different countries are carrying out laboratory tests to detect Covid-19 cases.\footnote{\url{https://ourworldindata.org/covid-testing}} The dataset on the number of tests carried out globally is published by {\it Our World in Data} in the following GitHub repository Owid/covid.19-data.\footnote{\url{https://github.com/owid/covid-19-data/tree/master/public/data/testing}}

\subsection{Examples of Regional Datasets}
The majority of the following datasets can be found in GitHub, searching by the term \textit{Covid-19}. 

\subsubsection{Africa}
For the African continent, reliable datasets can be found at \url{https://github.com/dsfsi/covid19africa} as GitHub repository \cite{marivate_vukosi_2020_3732980}.

\subsubsection{Argentina}
The Argentina Ministry of Health provides daily updates on the Covid-19 spreading, including data on the number of infected people divided by regions.\footnote{\url{https://www.argentina.gob.ar/coronavirus/informe-diario}}
The data can be downloaded from the GitHub repository \href{https://github.com/SistemasMapache/Covid19arData}{Covid19arData}.

\subsubsection{Australia}
The Health Department \href{https://www.health.gov.au/news/health-alerts/novel-coronavirus-2019-ncov-health-alert/coronavirus-covid-19-current-situation-and-case-numbers}{Health Department} of the Australian Government publishes the Covid-19 data. The data corresponding to the different regions of Australia can be downloaded from the JHU Github repository.\footnote{\url{https://github.com/CSSEGISandData/COVID-19/}} The regional Australian Covid-19 data are integrated into the global timeseries .csv files, which include information on confirmed cases, number of deaths, and number of recovered. See section \ref{SubSubSection:JHU:Data:Set} for further details. An additional GitHub repository is available at
\url{https://github.com/covid-19-au/covid-19-au.github.io}.
\subsubsection{China}
The data corresponding to the different regions of China can be downloaded from the JHU Github repository.\footnote{\url{https://github.com/CSSEGISandData/COVID-19/}} The regional Covid-19 Chinese data is integrated into the global time series .csv files. Moreover, the National Health Commission of the People's Republic of China updates daily the available information on the situation in China.\footnote{\url{http://en.nhc.gov.cn/DailyBriefing.html}}
Relevant information about the pandemic in China can also be found at the Midas GitHub repository: \url{https://github.com/midas-network/COVID-19/tree/master/data/cases/china}.

\subsubsection{France}
The data corresponding to France is provided by the different regions and published by the \href{https://www.santepubliquefrance.fr}{Public France Health System} at the official open data portal \url{https://www.data.gouv.fr/}. Among the different datasets available under the search of the term \textit{Covid}, three of them are highlighted by the portal (organized into .csv files):
\begin{itemize}
    \item \href{https://www.data.gouv.fr/fr/datasets/donnees-hospitalieres-relatives-a-lepidemie-de-covid-19/}{Covid-19 Hospital Data}: hospitalized cases, ICU cases, deaths per department, region, gender and age range \cite{France:Hospital:Data:Set}.
    \item \href{https://www.data.gouv.fr/en/datasets/donnees-des-urgences-hospitalieres-et-de-sos-medecins-relatives-a-lepidemie-de-covid-19/}{Covid-19 Emergency Room Admissions}: hospitalized cases, ICU cases, deaths per department, region, gender and age range \cite{France:Urgences:Data:Set}. 
    \item \href{https://www.data.gouv.fr/fr/datasets/donnees-relatives-aux-tests-de-depistage-de-covid-19-realises-en-laboratoire-de-ville/}{Covid-19 Laboratory Tests}: number of positive and negative laboratory tests per department, gender, and age group \cite{France:Tests:Data:Set}. \end{itemize}
\noindent
Moreover, French Covid-19 datasets can be found in these two GitHub repositories: i) \href{https://github.com/opencovid19-fr/data/blob/master/README.en.md}{Covid-19 epidemic French national data}; and ii) \href{https://github.com/cedricguadalupe/FRANCE-COVID-19}{Projet d'historisation du nombre de cas par région du Covid-19}.
Finally, the national mortality register can be accessed at \url{https://www.insee.fr/fr/information/4470857#tableau-figure1_radio1} to compare the number of deaths with previous years.

\subsubsection{Germany}
The main official open data provider in Germany is the Robert Koch Institute\footnote{\url{https://www.rki.de/EN/Home/homepage_node.html}.}, a public health institute in Germany. It provides, by means of a catalogue of infectious diseases,\footnote{\url{https://www.rki.de/D.E./Content/InfAZ/InfAZ_marginal_node.html}} pertinent information on each disease listed in the catalogue, e.g., SARS. In particular for Covid-19, data on risk assessments, spread of the epidemic, epidemiological studies, etc., can be found at \url{https://www.rki.de/D.E./Content/InfAZ/S/SARS/SARS.html?nn=2386228}. Moreover, it provides also daily reports of Covid-19 outbreak in Germany 
\cite{Germany:Situation:Reports}. Additional data on Covid-19 case numbers in Germany, divided by state over time, can be found at the GitHub repository \url{https://github.com/jgehrcke/covid-19-germany-gae}. The national mortality register can be found at \href{https://www.destatis.de/EN/Themes/Society-Environment/Population/Deaths-Life-Expectancy/_node.html}{Destatis}.

\subsubsection{Iceland}
During this pandemic, it has been reported\footnote{\url{https://ourworldindata.org/covid-testing}} that Iceland is one the best countries in terms of data on testing population. The Iceland government publishes all information at the following link:  \url{https://www.covid.is/data}. A GitHub repository can be found at \url{https://github.com/gaui/covid19}.

\subsubsection{Italy}
The \href{http://www.protezionecivile.gov.it/attivita-rischi/rischio-sanitario/emergenze/coronavirus}{Italian Civil Protection Department}, i.e., the national body in Italy that deals with the prediction, prevention and management of emergency events, daily updates a GitHub repository organized by regions and provinces, where the Covid-19 time-series can be downloaded  (\url{https://github.com/pcm-dpc/COVID-19}). 
The .csv file corresponding to the daily data of each of the 20 Italian regions\footnote{Available at \url{https://github.com/pcm-dpc/COVID-19/raw/master/dati-regioni/dpc-covid19-ita-regioni.csv}.} provides the number of confirmed cases, deaths, recovered, hospitalized, confined at home and ICU cases, in addition to the number of daily tests. Furthermore, GEDI \textit{Gruppo Editoriale}, a relevant Italian media conglomerate, provides a portal where those data are arranged in several interactive graphs, which include also the impact on the local mobility.\footnote{\url{https://lab.gedidigital.it/gedi-visual/2020/coronavirus-in-italia/}}
The national mortality \href{http://dati.istat.it/Index.aspx?QueryId=19670}{register} of Italy can be consulted in order to evaluate the magnitude of the epidemic with respect the number of deaths in previous years.

\subsubsection{Paraguay}
The official portal for data reports on Covid-19 for Paraguay can be found at \url{https://www.mspbs.gov.py/reporte-covid19.html}. The data are provided by the Public Health system and the reports are stratified by age and gender, including data about the number of cases, number of deaths, and recovered people. Furthermore, data on Covid-19 spreading in Paraguay can be also found at \url{https://github.com/torresmateo/covidpy-rest/blob/master/data/covidpy.csv} as GitHub repository.

\subsubsection{South Africa}
The information on Covid-19 spreading in South Africa can be found at \url{https://github.com/dsfsi/covid19za} \cite{marivate_vukosi_2020_3732419} \cite{marivate2020use} as GitHub repository. The repository, named 
\textit{Covid-19 Data for South Africa}, is maintained and hosted by Data Science for Social Impact research group, led by Dr Vukosi Marivate, at the University of Pretoria. These data have been used in \cite{marivate2020framework} to determine what data should be included in a public repository amidst the COVID-19 outbreak and how this data should be disseminated within a public dashboard.

\subsubsection{South Korea}
Korea Centers for Disease Control and Prevention (KCDC) provides data sets on Covid-19 cases regularly at \url{https://www.cdc.go.kr/board/board.es?mid=a30402000000&bid=0030}. 
A specific GitHub repository is available at \url{https://github.com/parksw3/COVID19-Korea}.

\subsubsection{Spain}
The regional Covid-19 Spanish data are collected by the Spanish government and they are available at the national open data portal \url{https://datos.gob.es/}. Different health datasets can be searched at its open data catalogue.\footnote{\url{https://datos.gob.es/es/catalogo?theme_id=salud}} The specific search \textit{Covid} provides datasets related to the global Spanish data classified into regions, e.g., \textit{Evolución de enfermedad por el coronavirus (Covid-19)}, or specific of a particular Spanish region, e.g., \textit{Evolución del coronavirus (Covid-19) en Euskadi}. In the GitHub repository \url{https://github.com/datadista/datasets/tree/master/COVID\%2019}, the Covid-19 timeseries by regions (CCAA) can be downloaded. Also, auxiliary information, like number of available ICUs per region before the pandemic outbreak, age distribution of confirmed cases, etc., can be found there. Furthermore, similar data can be also found at \url{https://www.epdata.es/} searching by the term \textit{Covid-19}. It is important to highlight that each of the different regions might report case numbers with different criteria. 
The national mortality register can be accessed at \href{https://www.isciii.es/QueHacemos/Servicios/VigilanciaSaludPublicaRENAVE/EnfermedadesTransmisibles/MoMo/Paginas/MoMo.aspx}{MoMo}.

\subsubsection{United Kingdom}
The UK government is collecting data and making them officially available by the Public Health England (PHE), i.e., the executive agency of the Department of Health and Social Care in the UK. The PHE took on the role of the Health Protection Agency, the National Treatment Agency for Substance Misuse and a number of other health bodies. The official open data resource provided by the UK government can be found at \url{https://www.gov.uk/government/publications/covid-19-track-coronavirus-cases}. This dashboard is showing reported cases by Upper Tier Local Authority in England (UTLA). An Excel file with relevant information can be downloaded from the dashboard. The information is organized at different levels: i) \textit{total number of confirmed cases and deaths in the UK}; ii) \textit{deaths by country}: England, Scotland, Wales and North Ireland; iii) \textit{deaths by NHS regions}: London, South East, South West, East of England, Midlands, North East and Yorkshire, North West; and iv) \textit{deaths by UTLA authorities}: daily cases at each of more than 149 different UTLAs. A description of how the confirmed and deaths cases are counted is also available at \url{https://www.gov.uk/guidance/coronavirus-covid-19-information-for-the-public#number-of-cases-and-deaths}. The .csv files corresponding to the number of confirmed cases and deaths can also be downloaded from the official public health system.\footnote{\url{https://www.gov.uk/government/publications/covid-19-track-coronavirus-cases}} Additional datasets reporting the UK Covid-19 cases can be found at  \url{https://github.com/tomwhite/covid-19-uk-data} as GitHub Repository. The national mortality register can be found at \href{https://www.ons.gov.uk/peoplepopulationandcommunity/birthsdeathsandmarriages/deaths/datasets/weeklyprovisionalfiguresondeathsregisteredinenglandandwales}{Office for National Statistics}.

\subsubsection{United States} 
The data corresponding to the United States can be obtained from the {\it 2019 Novel Coronavirus Covid-19 (2019-nCoV)} dataset from Johns Hopkins University Center for Systems Science and Engineering (JHU CSSE). This dataset is available as GitHub repository at \url{https://github.com/CSSEGISandData/COVID-19/}, which is daily updated by JHU-CSSS itself \cite{CSSE}. Another relevant source for the U.S. is the Centers for Disease Control and Prevention (CDC).\footnote{\url{https://www.cdc.gov/}}
This entity publishes different data on the Covid-19 cases by state and auxiliary information as the number of tests carried out. The CDC also publishes weekly surveillance reports, which can be found at \url{https://www.cdc.gov/coronavirus/2019-ncov/cases-updates/}. Moreover, the COVID Tracking Project\footnote{\url{https://covidtracking.com/data}} collects and publishes the testing data available for the US states and territories, divided by states. Similar information can be obtained from \url{https://coronavirus.1point3acres.com/en}, including Canada. Last, the New York Times is releasing a series of data files with cumulative counts of Covid-19 cases in the US, at state and county level, over time. These data can be found at \url{https://github.com/nytimes/covid-19-data} as Github repository. 
\vspace{0.5cm}
 
\begin{table}[ht]
\begin{tabular}{|
>{\columncolor[HTML]{DAE8FC}}c |
>{\columncolor[HTML]{67FD9A}}c |
>{\columncolor[HTML]{FFFC9E}}l |}
\hline
\multicolumn{1}{|l|}{\cellcolor[HTML]{FFCCC9}} & \cellcolor[HTML]{FFFC9E}\textbf{Source}                                                                        & \cellcolor[HTML]{ECF4FF} GitHub repositories \\ \hline
\textbf{Argentina}                             & 
\href{https://www.argentina.gob.ar/coronavirus/informe-diario}{Ministry of Health} &  
\href{https://github.com/SistemasMapache/Covid19arData}{Covid19arData}
\\ \hline
\textbf{Australia}                             & 
\href{https://www.health.gov.au/news/health-alerts/novel-coronavirus-2019-ncov-health-alert/coronavirus-covid-19-current-situation-and-case-numbers}{Australian Health Department} &    \href{https://github.com/covid-19-au/covid-19-au.github.io}{covid-19-au}       \\ \hline
\textbf{China}                                 & 
\href{http://en.nhc.gov.cn/DailyBriefing.html}{China National Health Commission} & 
\href{https://github.com/CSSEGISandData/COVID-19/}{JHU}, \href{https://github.com/midas-network/COVID-19/tree/master/data/cases/china}{Midas-China} \\ \hline
\textbf{France}                                & 
 \href{https://www.santepubliquefrance.fr}{Public France Health System} &
 \href{https://github.com/opencovid19-fr/data/blob/master/README.en.md}{opencovid19-fr}, \href{https://github.com/cedricguadalupe/FRANCE-COVID-19}{FRANCE-COVID-19}\\ \hline
\textbf{Germany}                               & 
\href{https://www.rki.de/EN/Home/homepage_node.html}{Robert Koch Institute} &     
\href{https://github.com/jgehrcke/covid-19-germany-gae}{covid-19-germany-gae}\\ \hline

\textbf{Iceland}                               & 
\href{https://www.covid.is/data}{Government of Iceland} & \href{https://github.com/gaui/covid19}{gaui-covid19}     
\\ \hline
\textbf{Italy}                                 & \href{http://www.protezionecivile.gov.it/attivita-rischi/rischio-sanitario/emergenze/coronavirus}{Italian Civil Protection Department}                                            &  
\href{https://github.com/pcm-dpc/COVID-19}{pcm-dpc} \\ \hline

\textbf{Paraguay}                              & 
 \href{https://www.mspbs.gov.py/reporte-covid19.html}{Ministry  of Public Health and Soc. Welfare} 
 &            
 \href{https://github.com/torresmateo/covidpy-rest/blob/master/data/covidpy.csv}{covidpy-rest}\\ \hline
\textbf{South Africa}                          &        
\href{https://www.nicd.ac.za/}{National Inst. Communicable Diseases}&  \href{https://github.com/dsfsi/covid19za}{covid19za}                               \\ \hline
\textbf{South Korea}                          &   
\href{https://www.cdc.go.kr/board/board.es?mid=a30402000000&bid=0030}{Centers for Disease Control and Prevention}&    
 \href{https://github.com/parksw3/COVID19-Korea}{COVID19-Korea} \\ \hline
\textbf{Spain}                          &   
\href{https://www.mscbs.gob.es/profesionales/saludPublica/ccayes/alertasActual/nCov-China/home.htm}{Ministry of Health}&    
\href{https://github.com/datadista/datasets/tree/master/COVID\%2019}{datadista-Covid-19} \\ \hline
 \textbf{United Kingdom}                        &    
 \href{https://www.gov.uk/government/publications/covid-19-track-coronavirus-cases}{Pubic Health England} &                 \href{https://github.com/tomwhite/covid-19-uk-data}{covid-19-uk-data}                 \\ \hline
\textbf{United States}                         &  
 \href{https://www.cdc.gov/}{Centers for Disease Control and Prevention} &   
  \href{https://github.com/CSSEGISandData/COVID-19/}{JHU}, \href{https://github.com/nytimes/covid-19-data}{Nytimes}\\ \hline
\end{tabular}
\caption{\label{tab:Regional Data} Some examples of regional Covid-19 data resources.}
\end{table}

%% file: Data_sets_of_relevant_variables_for_Covid_19_analysis.tex
\section{Data sets of relevant variables for Covid-19 analysis}\label{Sec:Relevant:Variables}
In this section, we include datasets relevant for the study and development of models of Covid-19, such as demography, government measures, weather, and climate data. These are variables that are under research to evaluate their influence on the virus propagation. 

\subsection{Demographics Datasets}
Demographics datasets are of significant importance for COVID-19 analysis. In this section, they have been arranged in three main groups: i) population; ii) population density; and iii) age structure. We highlight the following datasets on population:
\begin{itemize}
    \item \textit{European Countries population}: the dataset Eurostat \textit{Population on 1 January} dataset from the EU open data portal, available at \url{https://appsso.eurostat.ec.europa.eu/nui/show.do?dataset=demo_pjan&lang=en}, provides the population information per country at 2019.
    \item \textit{Global population}: the Population reference Bureau has published at \url{https://www.prb.org/worldpopdata/} 2019 World Population Data Sheet.
    \item \textit{List of countries by their population 2020}: the global population at 2020 per country can be retrieved on  Kaggle.\footnote{\url{https://www.kaggle.com/tanuprabhu/population-by-country-2020}} The dataset contains not only population values but also other features for each country.
    \end{itemize}
\noindent
Moreover, there are also some portals that provide demographic information, e.g. population pyramid as at \url{https://www.populationpyramid.net/}.

For the population density, the European Environment Agency provided the \textit{Population density dis-aggregated with Corine land cover 2000} dataset as a GeoTiFF format file, which can be found at \url{https://www.eea.europa.eu/data-and-maps/data/population-density-disaggregated-with-corine-land-cover-2000-2}.

Last, about age structure, {\it Our World in Data} provides a report on the present situation on the planet, divided by countries, \cite{owidagestructure}. The corresponding dataset is made available at \url{https://ourworldindata.org/age-structure}.

\subsection{Datasets on Government Measures }
For datasets related to the government measures, ACAPS publishes reports and datasets on government measures on Covid-19 at \url{https://www.acaps.org/projects/covid19} (see section \ref{sub:sec:ACAPS} for further deatils). In particular, updated reports can be downloaded from \cite{Acaps:GovernmentMeasures:Report}. Moreover, the ACAPS \#COVID19 Government Measures Dataset \cite{Acaps:GovernmentMeasures:DataSet} puts together the measures implemented by governments worldwide in response to the Covid-19 pandemic. The researched information available falls into five categories: i) social distancing; ii) movement restrictions; iii) Public Health measures; iv) social and economic measures; and v) lockdowns. Each category is broken down into several types of measures. 

Another source for government measures to fight Covid-19 outbreak is made available by the OxCGRT group at Oxford University, which provides an online API to access the country stringency data available at \url{https://covidtracker.bsg.ox.ac.uk/about-api} whereas the full dataset can be found at \url{https://www.bsg.ox.ac.uk/research/research-projects/oxford-covid-19-government-response-tracker}.

The EpidemicForecasting.org website provides a dataset on mitigation measures carried out by countries, which can be found at \url{http://epidemicforecasting.org/about}. In addition to that, they provide a simulator, i.e. the GLEAMviz simulator, which allows to explore realistic epidemic spreading scenarios at the global scale.\footnote{\url{http://www.gleamviz.org/simulator/}} 

An application named CHIME, i.e., COVID-19 Hospital Impact Model for Epidemics, has been developed by the Penn Medicine academic medical center from the University of Pennsylvania. This app is designed to assist hospitals and public health officials to understand hospital capacity needs as they relate to the COVID-19 pandemic. The application is based on a data model available at \url{https://code-for-philly.gitbook.io/chime/}. 

Last, the Open Government Partnership (OPG) organization has created a list of open government approaches to fight Covid-19 available at \url{https://www.opengovpartnership.org/collecting-open-government-approaches-to-covid-19/}. In particular, these approaches are organized by country and regions, and a brief description and related URL are also provided for each one. 

\subsection{Weather DataSets and Applications}
In this section, we focus on datasets related to weather, which are provided by several organizations all around the world, as described in the follows.

The first group of organizations providing weather datasets are the following EU providers:
\begin{itemize}
    \item {\it European Centre for Medium-Range Weather Forecasts (ECMWF)}: is a research institute and an operational service, producing global numerical weather predictions and other data. It operates two services from the EU’s Copernicus Earth observation programme, the Copernicus Atmosphere Monitoring Service (CAMS) and the Copernicus Climate Change Service (C3S). Two main services are provided by the ECMWF. The first one is the European Climate Data Store: The European Commission has entrusted ECMWF with the implementation of the Copernicus Climate Change Service (C3S). The mission of C3S is to provide authoritative, quality-assured information to support adaptation and mitigation policies in a changing climate. At the heart of the C3S infrastructure is the Climate Data Store (CDS),\footnote{\url{https://cds.climate.copernicus.eu/}} which provides information about the past, present and future climate in terms of Essential Climate Variables (ECVs) and derived climate indicators. The second ECMWF service is the Copernicus Climate Change Service (C3S*), which has worked with environmental software experts B-Open\footnote{\url{https://www.bopen.eu/}} to develop an application that allows health authorities and epidemiology centres to explore whether temperature and humidity affect the spread of the coronavirus. This application is freely accessible from the C3S Climate Data Store \cite{Copernicus:seasonal:behaviour}.
    \item {\it European Commission’s Joint Research Centre (JRC)}: different open-data projects at JRC can be of interest for the scientific community fighting Covid-19. We highlight here the most relevant one, represented by the Photovoltaic Geographical Information System (PVGIS). The focus of PVGIS is the research in solar resource assessment, photovoltaic (PV) performance studies, and the dissemination of knowledge and data about solar radiation and PV performance. The PVGIS web application\footnote{\url{https://ec.europa.eu/jrc/en/pvgis}} allows to access to meteorological data pertinent to the study of the seasonal behaviour of the pandemic. Three tools are available: i) Photovoltaic Performance; ii) Solar Radiation; and iii) Typical Meteorological Year (TMY tool).
\end{itemize}

The second group of organizations providing weather datasets are US providers: i) the National Oceanic and Atmospheric Administration (NOAA); and ii) the National Aeronautics and Space Administration (NASA). NOAA is an American scientific agency within the United States Department of Commerce that focuses on the conditions of the oceans, major waterways, and the atmosphere. It provides throught its open climate data portal\footnote{Climate Data Online (CDO): \url{https://www.ncdc.noaa.gov/cdo-web/}.} free access to global historical weather and climate data, in addition to station history information. These data include quality controlled daily, monthly, seasonal, and yearly measurements of temperature, precipitation, wind, etc.

On the other hand, NASA's goal in Earth science is to observe, understand, and model the Earth system to discover how it is changing. From an open-data perspective, NASA's project Prediction of Worldwide Energy Resource (POWER) can be very useful to recollect time series and monthly means of the most relevant weather and climate variables for a given location. POWER project\footnote{\url{https://power.larc.nasa.gov/}} was initiated to improve upon the current renewable energy data set and to create new data sets from new satellite systems. The POWER project targets three user communities: (1) Renewable Energy; (2) Sustainable Buildings; and (3) Agroclimatology. The access to the information can be done through the Data Access Viewer at \url{https://power.larc.nasa.gov/data-access-viewer/}, which is a responsive web mapping application providing data sub-setting, charting, and visualization tools in an easy-to-use interface. 


Last, there are many online APIs that provide weather data\footnote{See, for example, the list presented in \url{https://datarade.ai/data-categories/weather-data/overview}.}. Some of them can be used free of charge for a limited number of requests. As an example, see {\it World Weather online}.\footnote{\url{https://www.worldweatheronline.com/developer/api/historical-weather-api.aspx}}

\subsection{Mobility data sets}
This section includes datasets related to mobility of people. 
\begin{itemize}
 \item \textbf{Mobility reports:} Google has developed Covid-19 Community Mobility Reports, in which each report is broken down by location and displays the change in visits to places, like grocery stores and parks. The reports can be obtained by location at \url{https://www.google.com/covid19/mobility/}. As a result, a PDF document can be downloaded containing figures and trends. A similar tool has been developed by Apple and it can be found at \url{https://www.apple.com/covid19/mobility}. The reports can be obtained filtering by country. One important difference with respect to the Google app is that the raw data can be retrieved in the form of .csv files. Last, the GeoDS Lab (Department of Geography at University of Wisconsin-Madison) has developed a web application to identify mobility pattern changes in the U.S. \cite{Gao2020mapping}. The application can be accessed at  \url{https://geods.geography.wisc.edu/covid19/physical-distancing/}. 
\item \textbf{Aiport connectivity:} FLIRT is a tool that allows to get data about commercial flights. It shows direct flights from a selected location, and can simulate passengers taking multi-leg itineraries. The data can be downloaded in different formats (.csv, JSON, etc.) at \url{https://flirt.eha.io/}
\item \textbf{Contact tracing:}
Another important source of data related to mobility for modeling the pandemic is human behaviour inferred from wireless technologies, such as cell communications, WiFi and Bluetooth, among others. On this line, CRAWDAD is the Community Resource for Archiving Wireless Data At Dartmouth, a wireless network data resource for the research community. This repository contains wireless trace data from many contributing locations, and staff to develop better tools for collecting, anonymizing, and analyzing the data. The repository can be accessed at \url{http://crawdad.org/index.html} and it allows to filter the data, for instance,  Human Behavior Modeling and Opportunistic Connectivity, among other fields.   
\end{itemize}

%% file: Reusability_of_Open_Data_Sources.tex
\section{Reusability of Open Data sources} \label{sec:Reusability}
To maximize the value of the data sources about Covid-19, it is necessary that data sources are not only available but also have a set of characteristics that make them reusable. Due to the global affection of the pandemic, data sources are most of the cases coming from public institutions. These open government data should follow the eight principles of open data as reported in \cite{Lessig}. 

MELODA 5 \cite{Meloda5} is a metric to assess the reusability of an open data datasets. This metric considers 8 dimensions that affects the reusability of a dataset, which are listed hereafter:
\begin{enumerate}
      \item {\it Legal license}: assesses the legal rights given to the reusers of the dataset.
    \item {\it Technical format}: assesses the digital storage format in which the data is stored and released.
    \item {\it Access}: assesses the possibilities offered to reusers to interact with the dataset to retrieve the necessary set of data. 
    \item {\it Standardization}: assesses how popular and agreed are the fields composing the dataset and its description.
    \item {\it Geolocalization}: assesses the geographical content of the released data.
    \item {\it Updating frequency}: assesses the frequency of updating of the dataset.
    \item {\it Dissemination}: assesses the efforts and resources done by the publishing entity to makes popular the released datasets.
    \item {\it Prestige}: assesses the reputation of he publishing entity for the reusers of their data.\footnote{For Covid-19, this dimension cannot be set due to the novelty of the phenomenon.}
\end{enumerate}
According to these dimensions, the assessment of the main data sources mentioned in previous sections is reported in the following list:\footnote{In this list the prestige dimension has not been removed and therefore there are 6 points of difference with next table.}
\begin{itemize}
    \item {\it 2019 Novel Coronavirus COVID-19 (2019-nCoV) Data Repository by JHU CSSE}: 31
    \item {\it Our world in data. Coronavirus Source Data}: 34
    \item {\it Argentina}: 30
    \item {\it Australia}: 34
    \item {\it China}: 31
    \item {\it Italy}: 38
    \item {\it France}: 34
    \item {\it Germany}: 37
    \item {\it Paraguay}: 25
    \item {\it South Africa}: 40
    \item {\it Spain}: 37
    \item {\it United Kingdom}: 35   
    \item {\it United States}: 32   
\end{itemize}

\begin{table}[ht]
\centering
\begin{tabular}{|l|c|c|c|c|c|c|c|c|c|c|c|c|c|}
\hline
\rowcolor[HTML]{FFFC9E} 
\cellcolor[HTML]{FFCCC9}                   & \multicolumn{1}{l|}{\cellcolor[HTML]{FFFC9E}AR} & \multicolumn{1}{l|}{\cellcolor[HTML]{FFFC9E}AU} & \multicolumn{1}{l|}{\cellcolor[HTML]{FFFC9E}CN} & \multicolumn{1}{l|}{\cellcolor[HTML]{FFFC9E}DE} & \multicolumn{1}{l|}{\cellcolor[HTML]{FFFC9E}FR} & \multicolumn{1}{l|}{\cellcolor[HTML]{FFFC9E}GB} & \multicolumn{1}{l|}{\cellcolor[HTML]{FFFC9E}IT} & \multicolumn{1}{l|}{\cellcolor[HTML]{FFFC9E}JHU} & \multicolumn{1}{l|}{\cellcolor[HTML]{FFFC9E}OWID} & \multicolumn{1}{l|}{\cellcolor[HTML]{FFFC9E}PY} & \multicolumn{1}{l|}{\cellcolor[HTML]{FFFC9E}SP} & \multicolumn{1}{l|}{\cellcolor[HTML]{FFFC9E}US} & \multicolumn{1}{l|}{\cellcolor[HTML]{FFFC9E}ZA} \\ \hline
\rowcolor[HTML]{67FD9A} 
\cellcolor[HTML]{DAE8FC}License            & \textbf{6}                                      & \textbf{3}                                      & \textbf{3}                                      & \textbf{6}                                      & \textbf{6}                                      & \textbf{6}                                      & \textbf{6}                                      & \textbf{3}                                       & \textbf{6}                                        & \textbf{1}                                      & \textbf{6}                                      & \textbf{6}                                      & \textbf{6}                                      \\ \hline
\rowcolor[HTML]{67FD9A} 
\cellcolor[HTML]{DAE8FC}Technical Format   & \textbf{1}                                      & \textbf{3}                                      & \textbf{3}                                      & \textbf{6}                                      & \textbf{3}                                      & \textbf{6}                                      & \textbf{6}                                      & \textbf{3}                                       & \textbf{3}                                        & \textbf{1}                                      & \textbf{6}                                      & \textbf{3}                                      & \textbf{6}                                      \\ \hline
\rowcolor[HTML]{67FD9A} 
\cellcolor[HTML]{DAE8FC}Access             & \textbf{1}                                      & \textbf{1}                                      & \textbf{1}                                      & \textbf{3}                                      & \textbf{1}                                      & \textbf{1}                                      & \textbf{1}                                      & \textbf{1}                                       & \textbf{1}                                        & \textbf{1}                                      & \textbf{1}                                      & \textbf{1}                                      & \textbf{6}                                      \\ \hline
\rowcolor[HTML]{67FD9A} 
\cellcolor[HTML]{DAE8FC}Standarization     & \textbf{1}                                      & \textbf{3}                                      & \textbf{3}                                      & \textbf{1}                                      & \textbf{3}                                      & \textbf{1}                                      & \textbf{1}                                      & \textbf{3}                                       & \textbf{3}                                        & \textbf{1}                                      & \textbf{3}                                      & \textbf{1}                                      & \textbf{1}                                      \\ \hline
\rowcolor[HTML]{67FD9A} 
\cellcolor[HTML]{DAE8FC}Geolocalization    & \textbf{3}                                      & \textbf{3}                                      & \textbf{3}                                      & \textbf{3}                                      & \textbf{3}                                      & \textbf{3}                                      & \textbf{6}                                      & \textbf{3}                                       & \textbf{3}                                        & \textbf{3}                                      & \textbf{3}                                      & \textbf{3}                                      & \textbf{3}                                      \\ \hline
\rowcolor[HTML]{67FD9A} 
\cellcolor[HTML]{DAE8FC}Updating Frequency & \textbf{6}                                      & \textbf{6}                                      & \textbf{6}                                      & \textbf{6}                                      & \textbf{6}                                      & \textbf{6}                                      & \textbf{6}                                      & \textbf{6}                                       & \textbf{6}                                        & \textbf{6}                                      & \textbf{6}                                      & \textbf{6}                                      & \textbf{6}                                      \\ \hline
\rowcolor[HTML]{67FD9A} 
\cellcolor[HTML]{DAE8FC}Dissemination      & \textbf{6}                                      & \textbf{6}                                      & \textbf{6}                                      & \textbf{6}                                      & \textbf{6}                                      & \textbf{6}                                      & \textbf{6}                                      & \textbf{6}                                       & \textbf{6}                                        & \textbf{6}                                      & \textbf{6}                                      & \textbf{6}                                      & \textbf{6}                                      \\ \hline
\rowcolor[HTML]{FFCE93} 
\cellcolor[HTML]{FE996B}TOTAL              & {\color[HTML]{FE0000} \textbf{24}}              & {\color[HTML]{FE0000} \textbf{25}}              & {\color[HTML]{FE0000} \textbf{25}}              & {\color[HTML]{FE0000} \textbf{31}}              & {\color[HTML]{FE0000} \textbf{28}}              & {\color[HTML]{FE0000} \textbf{29}}              & {\color[HTML]{FE0000} \textbf{32}}              & {\color[HTML]{FE0000} \textbf{25}}               & {\color[HTML]{FE0000} \textbf{28}}                & {\color[HTML]{FE0000} \textbf{19}}              & {\color[HTML]{FE0000} \textbf{31}}              & {\color[HTML]{FE0000} \textbf{26}}              & {\color[HTML]{FE0000} \textbf{34}}              \\ \hline
\end{tabular}
\caption{\label{tab:Reusability} Total score corresponding to the first 7 reusability dimensions of MELODA 5 for different open institutional data sources. (AR: Argentina; AU: Australia; CN: China; DE: Germany; FR: France; GB: United Kingdom; IT: Italy; JHU: Johns Hopkins University; OWID: Our World In Data; PY: Paraguay;  SP: Spain; US: United States; ZA: South Africa).}
\end{table}

Although the maximum score for MELODA 5 is 61 points \cite{Meloda5}, in Table \ref{tab:Reusability} the \textit{Prestige} dimension of the publishing institution regarding Covid-19 is not included due to the novelty of the situation. To obtain a fair comparison, this criterion has been removed from the analysis, thus leaving only 7 dimensions for the data sources. Accordingly, a maximum of 55 points can be achieved. From this table, it is clear that none of the sources score results higher than 35 points, a value that can be considered good but far from optimum. We shall highlight that some sources are releasing their data with a license that restrict commercial use (they are not open data).\footnote{See definition of open data at \href{http://opendefinition.org}.} Hence, a score of 1 has been set for them on \textit{License} dimension. 
Another remarkable point is the general lack of an API to access individual data in the data sources. This forces the reusers to update the full dataset on a daily basis. For this reason, most of the data sources score 1 in \textit{Access} dimension. It is also remarkable the general lack of geolocalization contents for most of the data sources.  A mere indication of the region/area is the most common geographic content. Consequently, 3 is the more frequent score for \textit{Geolocalization} dimension. Regarding technical format, .csv is the most popular, together with some sources using JSON file formats. This last format provides additional key identification for each value. Although many sources include a definition of the field, no Standardization effort is detected for sharing the same information between sources. In fact, there is a myriad of different field names and contents.
Hence, the \textit{Dissemination} dimension score has been considered the maximum for those sources that have a website to disseminate the data sources.

%% file: Conclusions_journal.tex
\section{Conclusions}\label{sec:conclusions}

In this paper, we provide a review of relevant open data sources for better understanding the worldwide spread of the Covid-19. We enumerate the variables required to obtain consistent epidemiological and forecasting models. In particular, we focus not only on the specific Covid-19 timeseries but also on a set of auxiliary variables related to the study of its potential seasonal behaviour, the effect of age structure and prevalence of secondary health conditions in the mortality, the effectiveness of government actions, etc. 

We analyze the present situation of the available Covid-19 open data. Unfortunately, it is far from ideal because of a good number of issues like data inconsistency, changing criteria, a large diversity of sources, non-comparable metrics between countries, delays, etc. Despite the difficulties, the availability of open data resources on Covid-19 and related variables provides many opportunities to different communities. In particular, epidemiologists, data-driven researches, health care specialists, machine learning community, data scientists, etc. With the goal of facilitating these communities the access to the required open-sources, we identify the principal open data entities pertinent to the study of Covid-19. Furthermore, we enumerate different open datasets, and their corresponding repositories, related to Covid-19 cases at a worldwide scale, but also at a regional/local level. In addition, we provide specific information about the data resources for a selection of countries that have been selected because of the intensity with which the pandemic has impacted them, or for their relevance in the seasonal study of Covid-19, e.g., south-hemisphere countries. Finally, we provide other open resources that facilitate the incorporation of demographics, weather and climate variables, etc.

%% file: Acknowledgments.tex
\section*{Acknowledgments}\label{sec:acknowledgments}
The authors belong to the {\bf CONCO-Team} (CONtrol COvid19 Team) and would like to thank the rest of its members for their support\footnote{The composition and goals of CONCO-team can be found at \url{https://github.com/CONCO-Team/CONtrol-COvid19-TEAM/blob/master/Conco\_Team\_Members_Goals\_and\_Contributions.pdf}.}. In addition, we would like to thank other contributors, such as {\bf Mr. Nadir Bouchama} Researcher at Centre de Reserche Sur L'Information Scientifique et Technique (Algeria); {\bf Dr. Ejay Nsugbe } researcher at Collins Aerospace UK; {\bf Dr. Federica Garin} researcher at Gipsa-Lab, Grenoble, France; {\bf  Dr. Vukosi Marivate}, Senior Lecturer at Department of Computer Science, Republic of South Africa;  
{\bf Dr. Terrence Patrick McGarty}, Research Associate at Research Laboratory of Electronics, Massachusetts Institute of Technology, USA; 
{\bf Dr. Sriram Gubbi} medical doctor at National Institute of Diabetes and Digestive and Kidney Diseases (NIDDK), USA; {\bf Dr. Ram\'on B\'ejar}  Associate Professor at Department of Computer and Industrial Engineerings, University of Lleida, Spain, and {\bf Dr. Thomas Meunier}, Associate Research at Department of Physical Oceanography, Woods Hole Oceanographic Institution, USA. 